\documentclass[11pt]{article}
\usepackage[pdftex]{graphicx}
\usepackage[cmex10]{amsmath}
\usepackage{amsfonts}
\usepackage{stfloats}
\usepackage{fullpage}
\usepackage{color}
\usepackage{amsmath,amsfonts,amssymb,url,setspace}

\newcommand{\mufr}{\mu_{\mathrm{fr}}}
\newcommand{\hatmufr}{\hat{\mu}_{\mathrm{fr}}}
\newcommand{\mue}{\mu_{\mathrm{e}}}
\newcommand{\ke}{K_{\mathrm{e}}}
\newcommand{\rhoe}{\rho_{\mathrm{e}}}
\newcommand{\hatrhoe}{\hat{\rho}_{\mathrm{e}}}
\newcommand{\hatmue}{\hat{\mu}_{\mathrm{e}}}
\newcommand{\hatke}{\hat{K}_{\mathrm{e}}}

\newcommand{\kappafr}{\kappa_{\mathrm{fr}}}
\newcommand{\kappaf}{\kappa_{\mathrm{f}}}
\newcommand{\kappas}{\kappa_{\mathrm{s}}}
\newcommand{\hatkappafr}{\hat{\kappa}_{\mathrm{fr}}}
\newcommand{\hatkappas}{\hat{\kappa}_{\mathrm{s}}}
\newcommand{\rhos}{\rho_{\mathrm{s}}}
\newcommand{\rhof}{\rho_{\mathrm{f}}}
\newcommand{\hatrhos}{\hat{\rho}_{\mathrm{s}}}

\newcommand{\cpI}{c_p^{\mathrm{I}}}
\newcommand{\cpII}{c_p^{\mathrm{II}}}

\newcommand{\vs}{v_{\mathrm{s}}}
\newcommand{\us}{u_{\mathrm{s}}}
\begin{document}

\title{Estimation of groundwater storage from seismic data using deep learning}
  
\author{Timo L\"ahivaara${}^{a,}$, Alireza Malehmir${}^b$, Antti Pasanen${}^c$, Leo K\"arkk\"ainen${}^{d,e}$, \\Janne M.J. Huttunen${}^d$, Jan S. Hesthaven${}^f$}

\date{${}^a$Department of Applied Physics, University of Eastern Finland, Kuopio, Finland\\\smallskip
       ${}^b$Department of Earth Sciences, Uppsala University, Uppsala, Sweden\\\smallskip
      ${}^c$Geological Survey of Finland, Kuopio, Finland\\\smallskip
      ${}^d$Nokia Bell Labs, Espoo, Finland\\\smallskip
      ${}^e$Department of Electrical Engineering and Automation, Aalto University, Espoo, Finland\\\smallskip
      ${}^f$Computational Mathematics and Simulation Science, Ecole Polytechnique F\'ed\'erale de Lausanne, Lausanne, Switzerland}
  
\maketitle
\subsection*{{\bf Abstract}}

Convolutional neural networks can provide a potential framework to characterize groundwater storage from seismic data. Estimation of key components such as the amount of groundwater stored in an aquifer and delineate water-table level, from active-source seismic data are investigated in this study. The data to train, validate, and test the neural networks are obtained by solving wave propagation in a coupled poroviscoelastic-elastic media. A discontinuous Galerkin method is applied to model wave propagation whereas a deep convolutional neural network is used for the parameter estimation problem. In the numerical experiment, the primary unknowns are the amount of stored groundwater and water-table level, and are estimated, while the remaining parameters, assumed to be of less of interest, are marginalized in the convolutional neural networks-based solution. Results, obtained through synthetic data, illustrate the potential of deep learning methods to extract additional aquifer information from seismic data, which otherwise would be impossible based on a set of reflection seismic sections or velocity tomograms.\bigskip

\section{Introduction}

Groundwater is the world’s largest readily available freshwater resource \cite{fetter} and of great importance in both developed and developing countries. A detailed knowledge of the underground water storage (aquifer) properties and subsurface parameters are crucial in aquifer management, e.g. preventing waterlevel drawdown and planning aquifer protection. Traditional approaches for studying aquifers include geophysical surveys, followed by drilling and hydraulic test studies. The methodology presented in this paper can potentially make a significant difference both economically, by reducing the number of boreholes, and for data coverage, by transitioning from point data to continuous data and extraction of important parameters from seismic data. 

Groundwater aquifers are found in porous media such as gravel or sand, or within fractured bedrock or porous limestones. One potential method to characterize and monitor aquifers is to employ seismic data. In seismic surveys, the signals propagating through porous aquifers are typically generated by vibrators or man-made impacts (impulsive sources). Because the seismic wave field interacts with porous materials, the poroelastic signature of the aquifer can be captured in seismograms. Hence, recorded seismic signals can be used, with efficient numerical tools, to increase the knowledge of the groundwater reservoir state. Porosity, it's subsurface distribution, and the water-table are the factors that define the amount of groundwater stored in an aquifer. Recent progress in computational methods for seismic wave propagation has made it possible, in principle, to attempt the estimation of the key aquifer parameters \cite{bosch10, lahivaara14, lahivaara15}. It is well-known that the porosity not only influences seismic velocities but also reduces the seismic amplitude through scattering and absorption. Hence, porosity can be estimated from seismic data \cite{karim,lahivaara_malmo}.

Neural networks have been previously applied to estimate groundwater levels and aquifer parameters. For example, \cite{couli}, \cite{DALIAKOPOULOS2005229}, and \cite{Mohanty2010} use input such as temperature and well-based water level measurements to build a model for predicting groundwater levels. Furthermore, \cite{BALKHAIR2002118} and \cite{Karahan2008} employ neural networks to estimate aquifer transmissivity and storativity from an applicable well-based dataset.

In this work, we consider prediction of the groundwater stored in an aquifer and the water-table level using seismic data. We use a Gaussian process to generate samples from the potential scenarios for the aquifer for both the material model and the geometry. The prior information for the aquifer system is based on earlier studies and traditional ray-tracing based estimates \cite{artimo, maries16}. For each sample, we compute seismic responses from multiple source locations and couple simulations with deep learning for the prediction of water-table level and the amount of stored water. More specifically, the presented approach comprises of two main components:
\begin{enumerate}
\item[I:] Seismic wave propagation from the source to receivers, i.e., the {\it forward problem}, in coupled poroviscoelastic-elastic media is simulated using a discontinuous Galerkin (DG) and low-storage Runge-Kutta time stepping methods \cite{hesthaven_warburton_book, lahivaara11, ward15}. The DG method is a well-known high-order accurate numerical technique to numerically solve differential equations and has properties that makes it well-suited for wave simulations, see e.g. \cite{kaser06, puente08, wilcox10, modave16, mathieu18}. These properties include, for example, the straightforward handling of complex geometries and large discontinuities in the material parameters. In addition, the method has excellent parallelization properties. All of these are essential features for the numerical scheme to be used for complex wave problems. 

\item[II:] The {\it inverse problem} of estimating the amount of groundwater and water-table level is addressed by a neural network. Compared to conventional inversion techniques, neural networks have the advantage, that after the network has been trained, inferences can be carried out by using the network without a need to evaluate the wave propagation solver. This dramatically reduces the computational time. Furthermore, the neural network provides a straightforward approach to marginalize uninteresting physical model parameters in the inference. In the application considered here, the majority of the unknown physical parameters, required to define the wave propagation model, are not of interest. All unknown model parameters are taken into account while solving the forward problem but the neural network itself is trained to recover only the water-table level and the amount of stored groundwater.

In this study, we consider a convolutional neural network (CNN) \cite{lecun98, Bengio09, LeCun15, buduma17} in which the computational burden can be substantially reduced as compared to traditional multi-layer fully connected networks. The use of convolution operations provides a very efficient framework to extract the low- and high-level features from the input data. Convolutional neural networks have proven their potential to interpret data in various estimation problems. Recent studies include electrical impedance tomography \cite{Hamilton2017}, aerosol research \cite{joutsensaari18}, and ultrasound tomography \cite{lahivaara18}. In the context of seismic imaging, deep learning techniques have been studied, e.g. in \cite{arayapolo17, arayapolo18}. 
\end{enumerate}

\section{Model setup}\label{sec:problem-formulation}
 
\begin{figure*}
\centering
\includegraphics[width=0.99\textwidth]{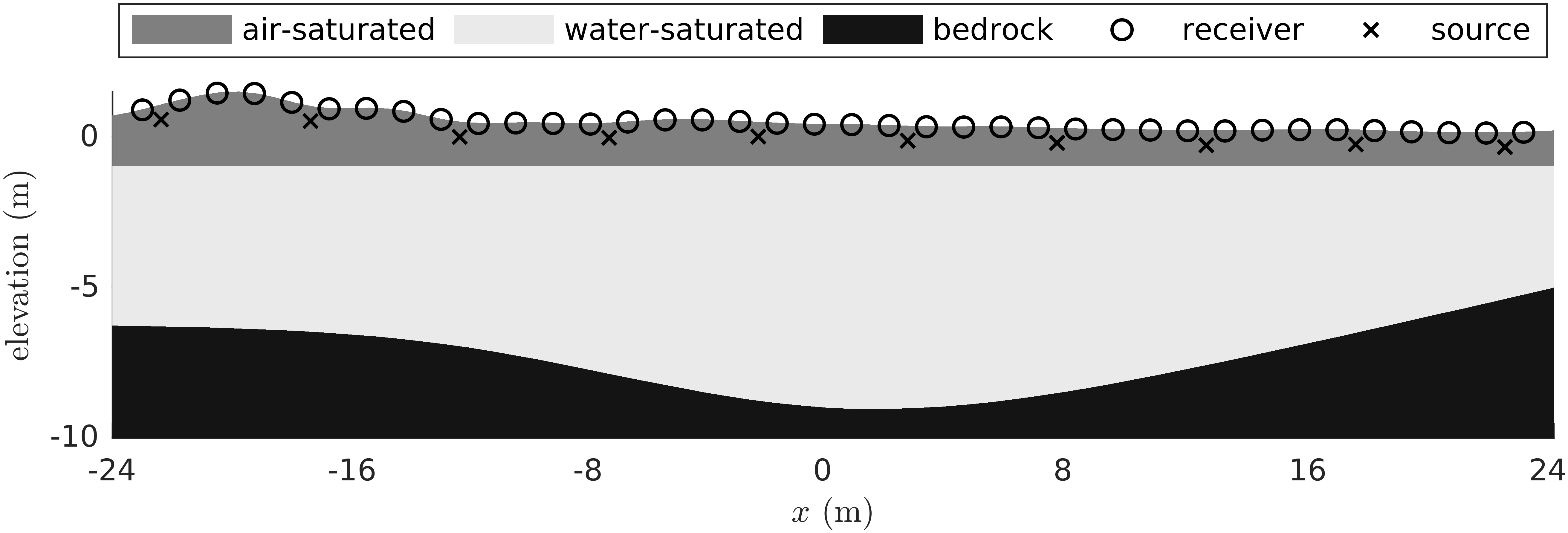}
\caption{\label{fig:geometry} Model used for benchmarking purposes. Circles denote the receivers and the crosses denote the sources.}
\end{figure*}

We consider a 48 m-long model domain $I=[-24,24]$ m, which consists of one elastic (bedrock) and two poroelastic (air- and water-saturated aquifer) subdomains, see Figure \ref{fig:geometry}. This model is motivated by an on-going seismic experiment carried out in an artificial water supply system located in Virttaankangas, in southwest Finland \cite{maries16}. The upper porous layer is air-saturated while the lower layer is water-saturated and the interface is the water-table. We asseme a free boundary condition on the top surface while other boundaries are modeled as outflow boundaries. Receivers are placed on the ground surface while sources are assumed to be buried at a depth of 0.5 m from the ground surface. In addition, 38 receivers measures the vertical velocity data from 10 seismic source positions.  The $x$-components of the receivers are distributed uniformly over the interval $x\in \left[-23,\ 23\right]$ m, while the $x$-components of the sources are distributed uniformly over the interval $x\in\left[-22.378,\ 22.378\right]$ m.

The seismic sources are modeled as force-type point (impulsive) sources, with the force pointing along the negative vertical-axis. We use the first derivative of a Gaussian pulse $\exp(-(\pi f_0(t-t_0))^2)$ as the time-dependent source signal with a frequency $f_0=100$ Hz and a time delay $t_0=1.2/f_0$.  The modeling time is 0.17 s. Note that recorded data were downsampled to a sampling frequency of 1 kHz on each receiver.

As the physical model, we use the coupled poroviscoelastic-elastic model studied in \cite{ward15}. The aquifer is modeled as a fully saturated porous material, based on the Biot theory, while the bedrock layer is assumed to have very low porosity and can therefore be modeled as an elastic layer. In the physical model for the poroviscoelastic media, a total of 11 physical parameters must be provided. These parameters are: fluid density $\rhof$, fluid bulk modulus $\kappaf$, viscosity $\eta$, solid density $\rhos$, solid bulk modulus $\kappas$, frame bulk modulus $\kappafr$, frame shear modulus $\mufr$, tortuosity $\tau$, porosity $\phi$, permeability $k$, and quality factor $Q_0$. In this study, we operate in the Biot's high-frequency regime for which the attenuation is controlled by the quality factor $Q_0$ \cite{carcione15, morency08, ward15}. In the elastic layer, we have a total of three unknown physical parameters, namely density $\rhoe$, bulk modulus $\ke$, and shear modulus $\mue$. For a more detailed discussion of the physical model, we refer to \cite{carcione15} and references therein.

\subsection*{Model parametrization}

The basement profile is taken from a ray-tracing based estimate $\hat{b}_y(x)$ \cite{maries16} to model parametrization and smoothing constraints used for fast model convergence. However, the basement profile is not assumed to be accurate. Instead, we consider profiles of the bedrock surface that are perturbed randomly. Basement profiles are sampled as 
\begin{equation}
  \label{eq:basement}
  b_y(x) = \hat{b}_y(x) + \delta_M M_c(x)+ \delta_H H_{x_H}(x),\quad x\in I,
\end{equation}
where $M_c$ is a Matern field \cite{rasmussen} with $\nu=3/2$ or, more specifically, a Gaussian process with the covariance function 
\begin{eqnarray}
  C_{\nu=3/2,c}(x,x')&=&\textrm{cov}(M_c(x),M_c(x'))\nonumber \\
  \label{eq:basement_C}
  &=&\left(1+\sqrt{3}\frac{\vert x-x'\vert} c\right)\exp\left(-\sqrt{3}\frac{\vert x-x'\vert} c\right), \quad x,x'\in I,
\end{eqnarray}
where $c$ is the characteristic length and $H$ is a Heaviside function ($H_{x_H}(x)=1$ if $x\geq x_H$ and 0 otherwise), used to produce a possible discontinuity in the basement surface profile.  

\begin{figure*}
\centering
 \includegraphics[width=0.99\textwidth]{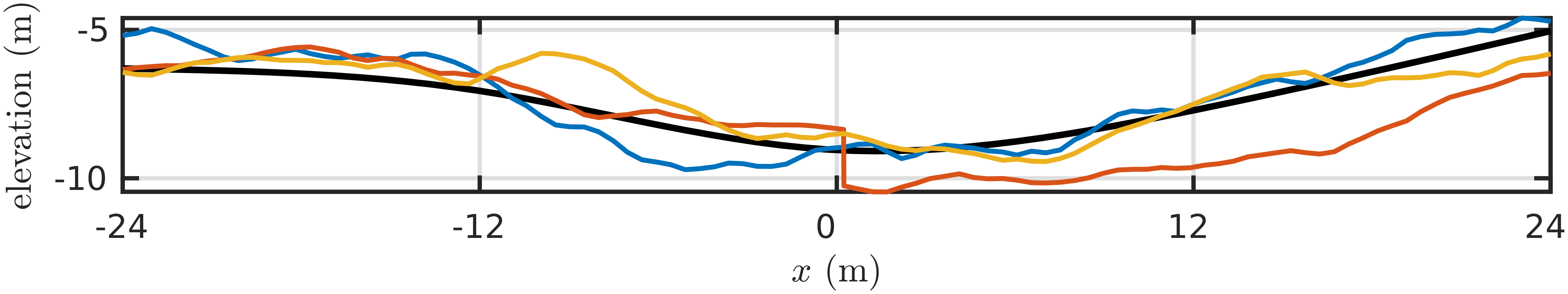}
\caption{\label{fig:basement}Three realizations and the ray-tracing based one (thick black line) to represent the basement surface profile.}
\end{figure*}

To generate a sample of a basement profile, we first sample the characteristic length $c$ and the standard deviation $\delta_M$ as $c \sim \mathcal{U}(3,\ 10)$ and $\delta_M \sim \mathcal{U}(0, 1)$, respectively, and then generate a realization of the Matern field corresponding to $c$. In addition, we assume that $\delta_H\sim \mathcal{U}(-2,2)$ and $x_H\sim \mathcal{U}(-24, 120)$, i.e., a discontinuity exists with probability $48/144=1/3$. Figure \ref{fig:basement} shows examples of the sampled basement surface profiles. The water-table level $W^l$ is assumed to vary uniformly between -3.7 m and -0.7 m whereas the ground surface remains fixed in each sample.

The fluid parameters for the water-saturated zone are given by: the density $\rhof=1000$ kg/m$^{3}$, the fluid bulk modulus $\kappaf=2.1025$ GPa, and the viscosity $\eta=1.3$e-3 Pa$\cdot$s, while in the air-saturated part, we set: $\rhof=1.2$ kg/m$^{3}$, $\kappaf=1.3628$e5 Pa, and $\eta=1.8$e-5 Pa$\cdot$s. All other material parameters of the aquifer are assumed to be unknown. Furthermore, the remaining material parameters are generated using an isotropic Ornstein-Uhlenbeck process $\Pi_c$ with the characteristic length $c$. Each unknown material parameter $\theta = \{\rhos,\ \kappas,\ \kappafr,\ \mufr,\ \tau,\ \phi,\ k,\ Q_0\}$ is modelled as follows:
\begin{eqnarray}
  \label{eq:param}
  &&\theta(x,y)  =  \theta^{\ast} + \delta_{\theta}\Pi_c(x, y)\\
  &&\theta^{\ast}  \sim  \mathcal{U}(0.9\hat{\theta},\
  1.1\hat{\theta}),\ \delta_{\theta} \sim \mathcal{U}(0,\
  0.1\theta^{\ast}),\ c \sim \mathcal{U}(2,\ 20)
\end{eqnarray}
Mean values of $\hat{\theta}$ are given in Table \ref{tab:bounds}. The unknown parameters are assumed to be uncorrelated.

\begin{table}
  \caption{The mean values of the uniform sampling distributions for each unknown physical parameter.  }\label{tab:bounds}
        \centering
  \begin{tabular}{cc|c}
  \hline
    \textbf{Variable name} & \textbf{Symbol}  & \textbf{Value}\\
    \hline
solid density&$\hatrhos$ (kg/m$^{3}$) & 2400\\
solid bulk modulus&$\hatkappas$ (GPa) & 3.0\\
frame bulk modulus&$\hatkappafr$ (GPa) & 0.3\\
frame shear modulus&$\hatmufr$ (GPa) & 0.2\\
tortuosity&$\hat{\tau}$  & 1.8\\
porosity&$\hat{\phi}$ (\%) & 30\\
permeability&$\hat{k}$ (m$^2$) & 5e-8\\
quality factor&$\hat{Q}_0$ & 50\\
\hline
    \end{tabular}
\end{table} 

In the bedrock layer, material fields are assumed to be purely elastic and homogeneous. Most of the energy reflects back from the bedrock surface and we expect the heterogeneity within the bedrock layer to have a minor effect. Therefore parameters $\theta=\{\ke,\ \mue,\ \rhoe\}$  are assumed to be homogeneous and are sampled as in (\ref{eq:param}) with $\delta_{\theta}=0$. For the mean values we set: $\hatke=66.0$ GPa, $\hatmue=24.75$ GPa, and $\hatrhoe=2750$ kg/m$^{3}$.

Table \ref{tab:derived_hete2} lists the calculated wave speeds for each subdomain. The reported wave speed values correspond to the values generated by sampling of the material parameters. For further details of calculating wave speeds, we refer to \cite{ward15}.

\begin{table}
  \caption{Calculated wave speeds for each subdomain. $\cpI$ and $\cpII$ denote the fast and slow pressure and $c_s$ the shear wave speed, respectively. Both minimum/maximum values, which are based on sampling the material parameters, are given.}\label{tab:derived_hete2}
        \centering
      \begin{tabular}{c|ccc}
      \hline
        {\bf Subdomain}  &  $\cpI$ (m/s)& $\cpII$ (m/s)& $c_s$ (m/s)\\
        \hline
        {\bf Air-saturated}   & 428/829 & 205/336 & 239/514\\
        {\bf Water-saturated} & 983/1563 & 210/470 & 231/478\\
        {\bf Bedrock}         & 5434/6598 & - & 2715/3312\\
      \hline
      \end{tabular}
\end{table}

The total amount of water stored in the aquifer is one of the primary unknowns in the application and is computed as
\begin{equation}
  \label{eq:water}
  V^w = \sum_{\ell=1}^{K^w}\phi^w_{\ell}A^w_{\ell}.
\end{equation}
Here, $K^w$ denotes the number of elements that belong to the water-saturated aquifer and $\phi^w_{\ell}$ and $A^w_{\ell}$ are the porosity and area of the $\ell$th element in the water-saturated aquifer, respectively. 

\begin{figure*}
\includegraphics[width=\textwidth]{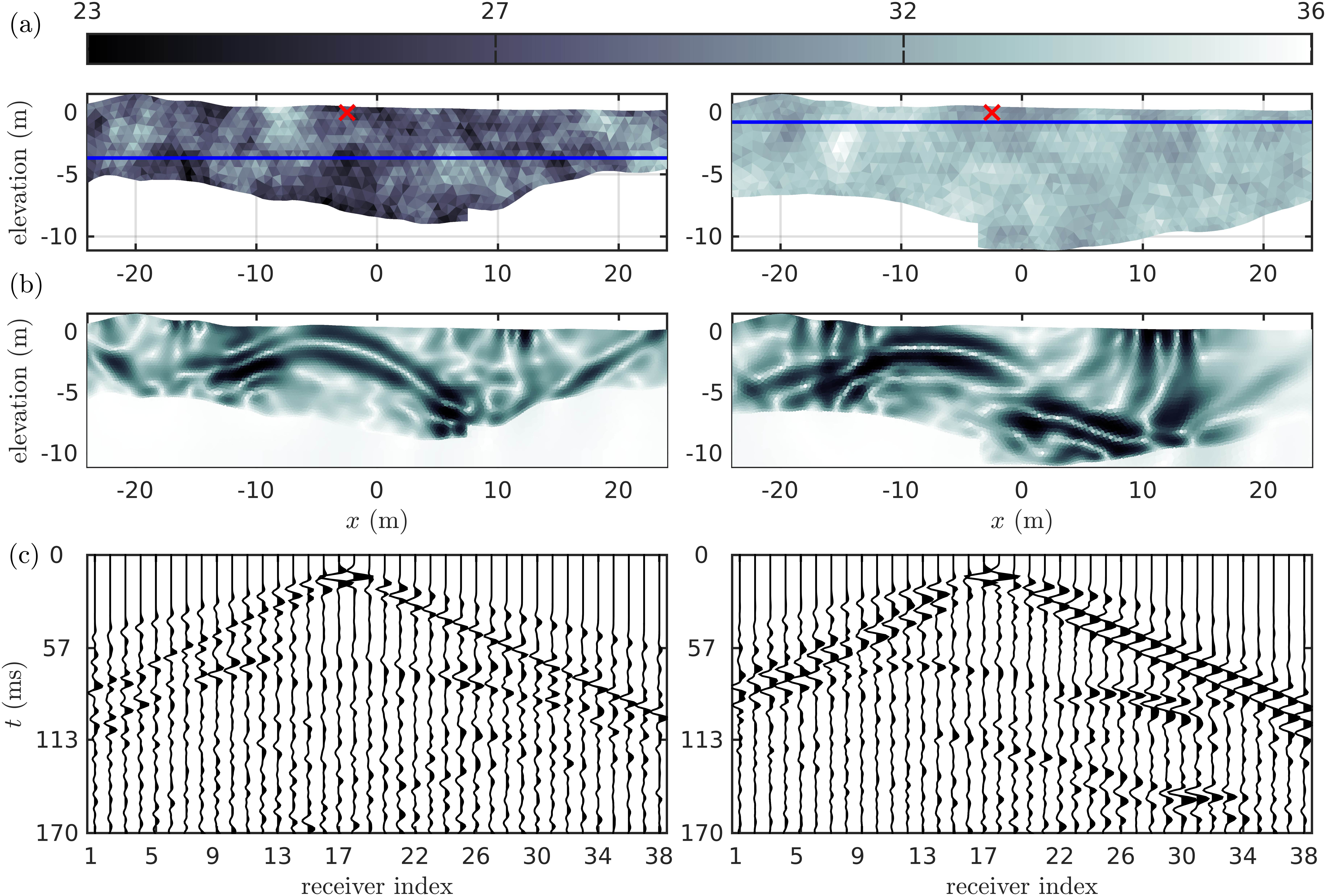}
\caption{Two example porosity fields (a) and the corresponding recovered vertical solid velocity data (seismogram) (c). Colorbar on top shows the porosity values in percentage. Snapshots in the middle (b) show the total solid velocity field for time instant 59 ms. Images correspond to samples where the amount of stored water is low (left) or high (right). The source location (red '$\times$') and water-table level (horizontal blue line) are shown on the top figures.  }  \label{fig:signals_teaching}
\end{figure*}

Two representative porosity fields are shown in Figure \ref{fig:signals_teaching}a.  Fields are shown for cases where the amount of stored groundwater is low (left) or high (right). The blue horizontal line shows the water-table level. The visualized porosity fields share the same color scale.

The snapshots of the solid velocity field $\sqrt{\us^2 + \vs^2}$, where $\us$ is the simulated horizontal and $\vs$ vertical component, are shown in the Figure \ref{fig:signals_teaching}b. The source location is shown in the top row. In contrast to the porosity visualization, the color scale is unique for the velocity fields. Snapshots are visualized at time 59 ms.  The velocity fields support the assumption that very low levels of energy penetrate to the elastic subdomain. Furthermore, the selected samples contain a sharp discontinuity in the basement profile, which can also clearly be seen from the snapshots. 

In Figure \ref{fig:signals_teaching}c, the corresponding noise-free shot records, i.e., the vertical solid velocity component $\vs$, are shown. As with the snapshots, the source location is given in the top row figures. The results show the direct arrival and a clear reflection from the basement boundary. The water-table produces no clear reflection as expected from the two heterogeneous models. Weak reflections from the boundaries are evident in the record and the wave field snapshot, particularly for the case of a deeper water-table.

\section{Deep convolutional neural networks}\label{sec:neural}

\begin{figure*}
\includegraphics[width=\textwidth]{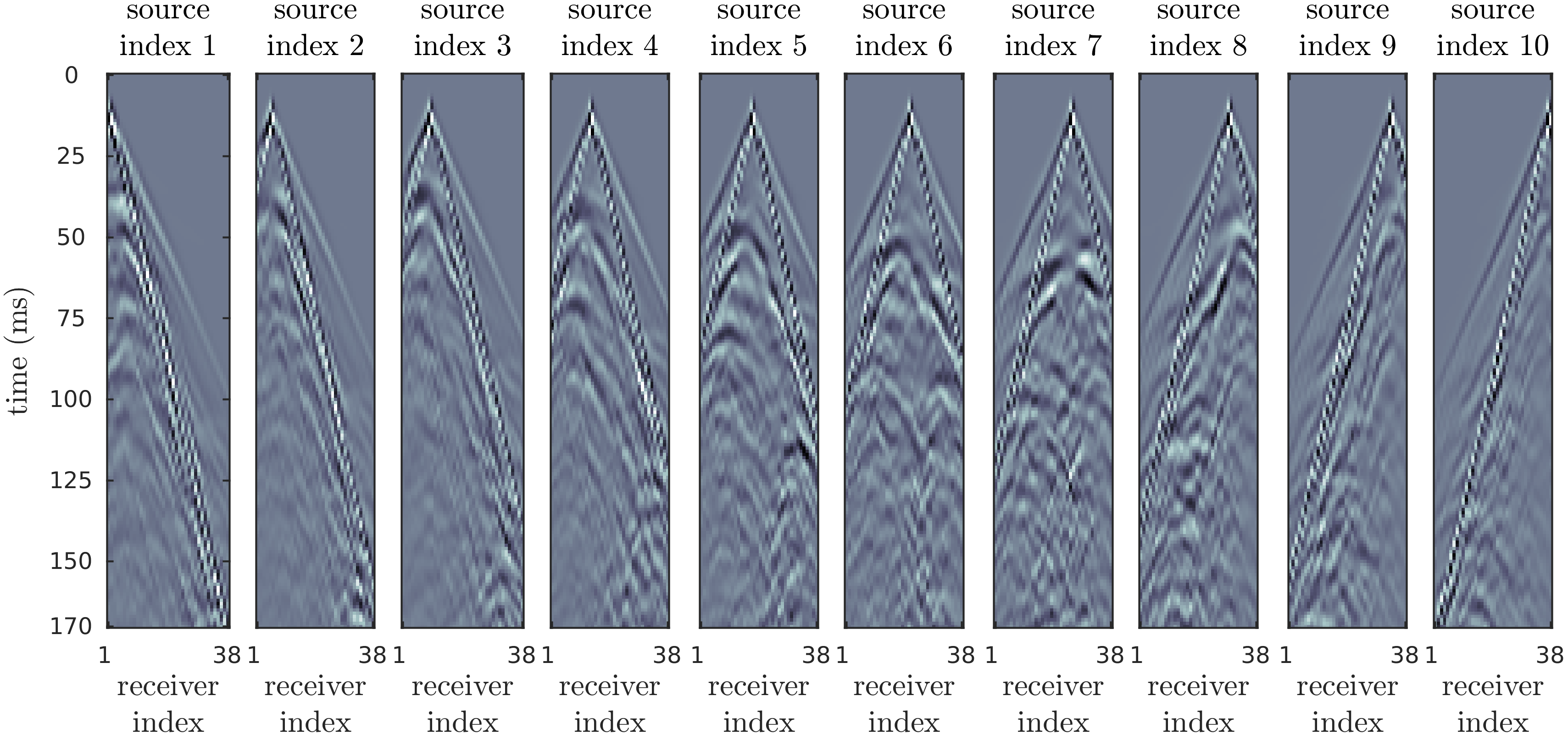}
\caption{A set of shot gathers, expressed as a stack of 2D-shot records obtained along the model shown in Figure \ref{fig:geometry}, used for training the deep convolutional neural networks. Note that the deep learning algorithm uses the original pixel values of the image $X$.} \label{fig:signals_teaching_1}
\end{figure*}

In this work, we apply a deep convolutional neural network to estimate the parameters related to groundwater storage. We consider a supervised regression task, i.e., the problem is to find a function $g$ from inputs $X$ and real values $\Theta$ (i.e. $\Theta = g(X)$) based on a set of input-output pairs $(X_\ell,\ \Theta_\ell)$, $\ell=1,\ldots, N_{nn},$ where $N_{nn}$ is the number of samples in the dataset, the generation of which is described below. In our problem, the input $X$ is the recorded vertical solid velocity wave data, expressed as 3D-data $X\in \mathbb{R} ^{d},\ d=N_t \times N_r \times N_s$, where $N_t$ denotes the number of time steps, $N_r$ the number of receivers, and $N_s$ the number of sources. In our case, the input dimension is $d=171 \times 38 \times 10$. An example of the velocity data are shown in Figure \ref{fig:signals_teaching_1}. We are interested in the amount of water stored in the aquifer $V^w$ (\ref{eq:water}) and the water-table level $W^l$, hence $\Theta$ is either $V^w$ or $W^l$.

The regression problem is often solved by expressing $g$ as a parameterized function $g_\vartheta$ such that $\Theta_\ell \approx g_\vartheta(X_\ell)$ for the samples in the training set. 
In machine learning, the function is often chosen to be a neural network \cite{buduma17}.   
For example, a neural network with two layers can be expressed as
\begin{eqnarray}
    A_1 & = & \sigma_1(w_1  X+b_1),\\
    g_\vartheta(X) & =& \sigma_2(w_2  A_1 + b_2),
\end{eqnarray}
where  $w_1$ and $w_2$ are matrices of weights, $b_1$ and $b_2$ are bias terms, and $\sigma_1$ and $\sigma_2$ are non-linear functions inducing nonlinearity to the network. 
Traditionally nonlinear functions were usually chosen to be tanh or sigmoid function, but lately the Rectified Linear Unit (ReLU), $\sigma=\max(0,X)$, has been widely used \cite{LeCun15}. 
The unknown parameters are $\vartheta=(w_1,w_2,b_1,b_2)$.

Layers employing the matrix multiplication (such as above) are called as fully connected (FC) layers. 
A disadvantage of FC layers is that the number of unknown weights can be very large especially if the input dimension is large.
To overcome this issue in image processing tasks, convolutional neural network (CNN) were introduced \cite{lecun98}.
In convolutional layers, the linear operation is chosen to be discrete convolution: for example,
\begin{equation}
A_1  =  \sigma_1(w_1\ast  X+b_1),\\
\end{equation}
where $\ast$ denotes the discrete convolution and $w_1$ is now a parametrized convolution kernel.
As a linear operation, the convolution can also be expressed as a matrix multiplication, but significant amount of elements of the corresponding matrix are either tied together or zeros, as a result the number of freedoms is significantly smaller.
In addition to the convolutions, CNN networks often also applies pooling layers that are used to reduce the dimensions of layer outputs.
For a detailed discussion, we refer to \cite{buduma17}.

\begin{figure*}
\includegraphics[width=\textwidth]{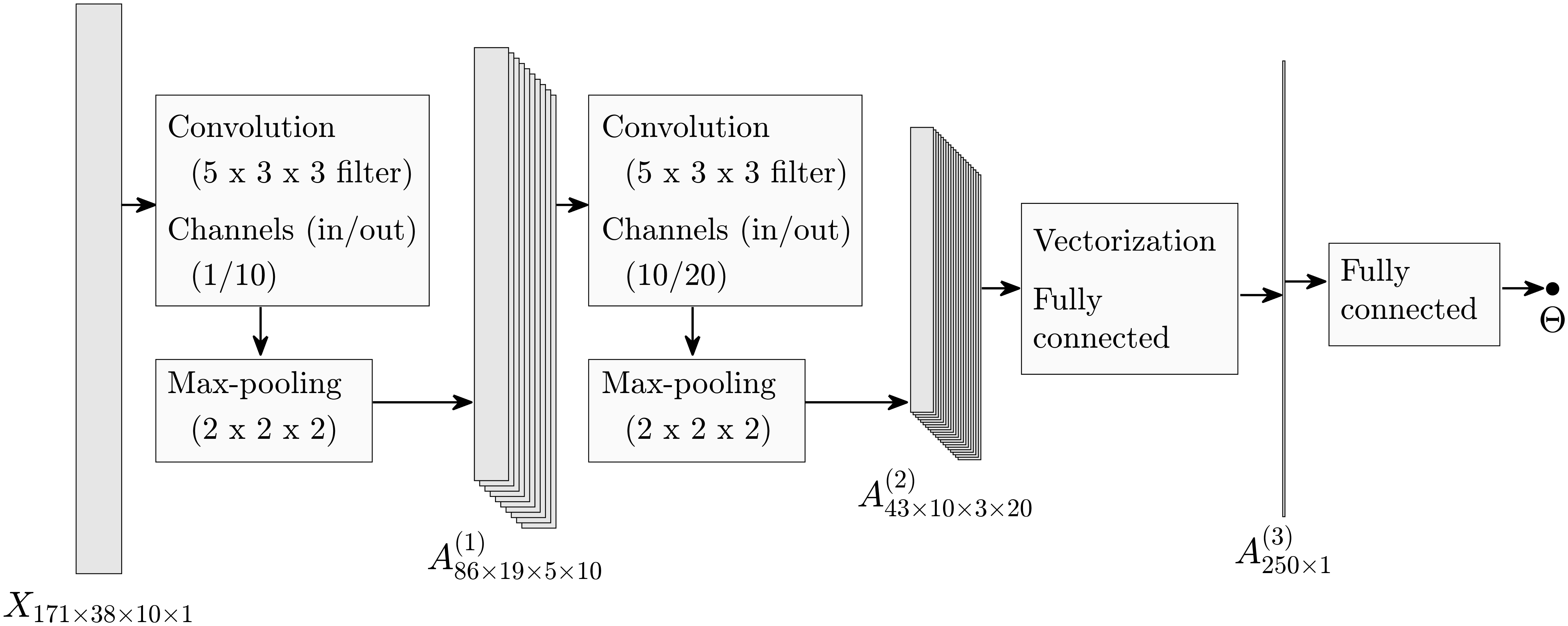}
\caption{Representation of the network architecture used in this work.} \label{fig:network}
\end{figure*}

The neural network used in our study comprises of two convolutional layers with max-pooling and two FC layers.
The details of the architecture are given in Table \ref{tab:network} and the architecture is also pictorially presented in Figure \ref{fig:network}.
The CNN architecture is similar to the ones used for image classification (e.g. ALEXNET \cite{krichevsky}), but with fewer convolutional layers and, instead of using softmax layer for classification, the last layer is chosen to be a linear layer for regression.
We have employed 3D convolutions as the input data is three-dimensional. 
The last dimension in the input typically represents the number of channels in the data (e.g. 3 for RGB images).
In our case, the last channel is chosen to be 1 as we have scalar valued data, i.e., the vertical solid velocity wave recordings; see Figure \ref{fig:signals_teaching_1}.

For regression problems, training of the model, i.e.,  finding parameters $\vartheta$, is typically carried out by minimizing the quadratic loss function $f(\vartheta;\{X_\ell, \Theta_\ell\})$ over the training set
\begin{equation}
  \label{eq:loss}
f(\vartheta;\{X_\ell, \Theta_\ell\} )=\frac{1}{N_{nn}}\sum_{\ell=1}^{N_{nn}} {
  (g_\vartheta(X_\ell ) - \Theta_\ell )^2}
\end{equation}
to obtain the weights and the biases of the network.
The optimization problem is typically solved using stochastic gradient descent method, in which the above sum is approximated with the sum calculated over randomly chosen minipatches, or some variants of this. 
We employ the Adam optimizer \cite{Kingma2014} for the optimization and the batch size is set to be 50 samples. For network training, a total of 200 full training cycles in stochastic optimization (epochs) was run. As the computing interface, a TensorFlow \cite{tensorflow2015} was used.

 \begin{table*}
  \caption{The convolutional neural network architecture used in this study.}\label{tab:network}
        \centering
    \begin{tabular}{c|c|c|c}
    \hline
    \textbf{Layer k} &  \textbf{Type and non-linearity} & \textbf{Input size} & \textbf{Output size}\\
    \hline
    & Input &  $171 \times 38\times 10\times 1$ &  \\    
    \hline
    1 & Convolution layer ($5\times 3\times 3$ filter) &  $171 \times 38\times 10\times 1$& \\
     &  + ReLU + Max-pooling ($2\times 2\times 2$)& & $86\times 19\times 5 \times 10$\\
    2 & Convolution layer ($5\times 3\times 3$ filter)&  $86\times 19\times 5 \times 10$& \\
     &  + ReLU + Max-pooling ($2\times 2\times 2$)& & $43\times 10\times 3 \times 20$\\
    3 & Vectorization          &  $43\times 10\times 3 \times 20$ & 25800 \\
      & Fully connected layer + ReLU          &  25800 & 250\\
    4 & Fully connected layer  &  250 & 1  \\
    \hline
                     & Output& &  1   \\
    \hline
    \end{tabular}
\end{table*}

Finally, we note that we have carried out experiments with different network architectures. For instance, increasing the number of layers an/or neurons did not improve the performance significantly. However, there was some performance decrease with a smaller number of layers or neurons. Selection for the filter size is based on testing different configurations, from which the best performing combination is chosen.  The chosen network architectures yields a good performance with the validation set, with a relatively small number of unknowns. We also carried out experiments using 2D convolutions, for example, to express the shot gather data as a single 2D image (stacking sensor measurements horizontally; see Figure \ref{fig:signals_teaching_1}) but this lead to inferior performance compared to 3D convolutions. 
However, our aim was not to find the most optimal network architecture and there may be other architectures giving a similar performance with even smaller number of the unknowns.

\subsection*{Training, validation, and test datasets}

For the training of the convolutional neural network, we generated a dataset comprising 15,000 samples, using computational grids that had $\sim$3.5 elements per wavelength. To monitor the network's generalization capabilities, we generated an additional validation dataset of 3,000 samples. The physical parameters and the geometry for each sample were drawn using the approach discussed earlier. To further control the numerical accuracy of the forward solver, the order $N_{\ell}$ of the polynomial basis approximation in each cell was allowed to vary in the elements of the computational grid. The order of the basis function in each element is selected from
\begin{equation}
\label{eq:porder}
  N_{\ell} = \left\lceil \frac{2\pi \bar{a}h^{\ell}_{\max}}{\lambda^w_{\ell}}+\bar{b}\right\rceil ,
\end{equation}
where  $\lceil\cdot\rceil$ is the ceiling function, $h^{\ell}_{\max}$ is the largest distance between two vertices, $\lambda^w_{\ell}=c^{\ell}_{\min}/f_0$ is the wavelength, $c^{\ell}_{\min}$ is the minimum wave speed, and parameters $\bar{a}$ and $\bar{b}$ control the local accuracy on each element.  For the generation of training and validation data, we set $(\bar{a},\ \bar{b})=(1.0294,\ 0.7857)$. We refer to \cite{lahivaara11, lahivaara10} for further details of the nonuniform basis orders. 

To train the model for robustness to the presence of measurement noise, the samples were corrupted with Gaussian noise. More precisely, we created five copies of each image in the dataset, which were then corrupted as
\begin{equation}
\label{eq:noisemodel}  
X_{\ell}^\textrm{noised}=X_\ell+A\alpha\epsilon^A+B|X_\ell|\epsilon^B,
\end{equation}
where $\alpha$ is the maximum absolute value of the training dataset and $\epsilon^{A/B}$ are independent zero-mean Gaussian random variables.  The second term represents additive (stationary) white noise and the last term represents noise relative to the signal amplitude.  To include a wide range of noise levels, the coefficients $A$ and $B$ for each sample were randomly chosen such that the standard deviations of the white noise component varies logarithmically between $(0.03-5)\%\alpha$, and the standard deviations of the relative component is between $(0-5)\%|X_\ell|$.  The total number of samples in the training set is $N_{nn}=5\times 15000=75000$.

To evaluate performance of the prediction algorithm, we generated a set of data (test set) similarly as the training set. For the test set, the computational grids were required to have $\sim$4 elements per wavelength and the non-uniform basis order parameters are $(\bar{a},\ \bar{b})=(1.2768,\ 1.4384)$ \cite{lahivaara11} of model (\ref{eq:porder}). The main reason to use denser discretization was to avoid an inverse crime \cite{kaipio07} related to simulation studies: the use of the same computational model, e.g. same discretizations, to generate both training and test data could potentially lead to situations in which severe modelling errors are ignored, resulting in unrealistic impressions of the accuracy of  the estimates when compared to the actual performance with real data. In the test data, noise was added in a more systematic manner to study the performance with different noise levels.

In the above test set, the ranges of material parameters were chosen to be same as in the generation of the training set. Typically it is a common practice to test performance using different distribution of data. However, the changing ranges of the model parameters may not be meaningful in such simulation study. Namely, testing performance ranges that are wider than in the training set can lead to poor performance as the neural network has to extrapolate outside of the data manifold (which can be rather random as the CNN model does not have any additional physical knowledge of the phenomenon). On the other hand, limited ranges would only correspond to a subset of the above test set. Nonetheless, to study the sensitivity of the network to incorrect model assumptions in the data generation, we generated another test set comprising of 200 samples that contain two jumps in the basement profile. These additional samples are generated with a modified version of the model (\ref{eq:basement}) by forcing the profile to have two jumps. For both jumps, the height is randomized as $\mathcal{U}(-2,2)$.

\section{Results}\label{sec:results}

\begin{figure*}
\includegraphics[width=\textwidth]{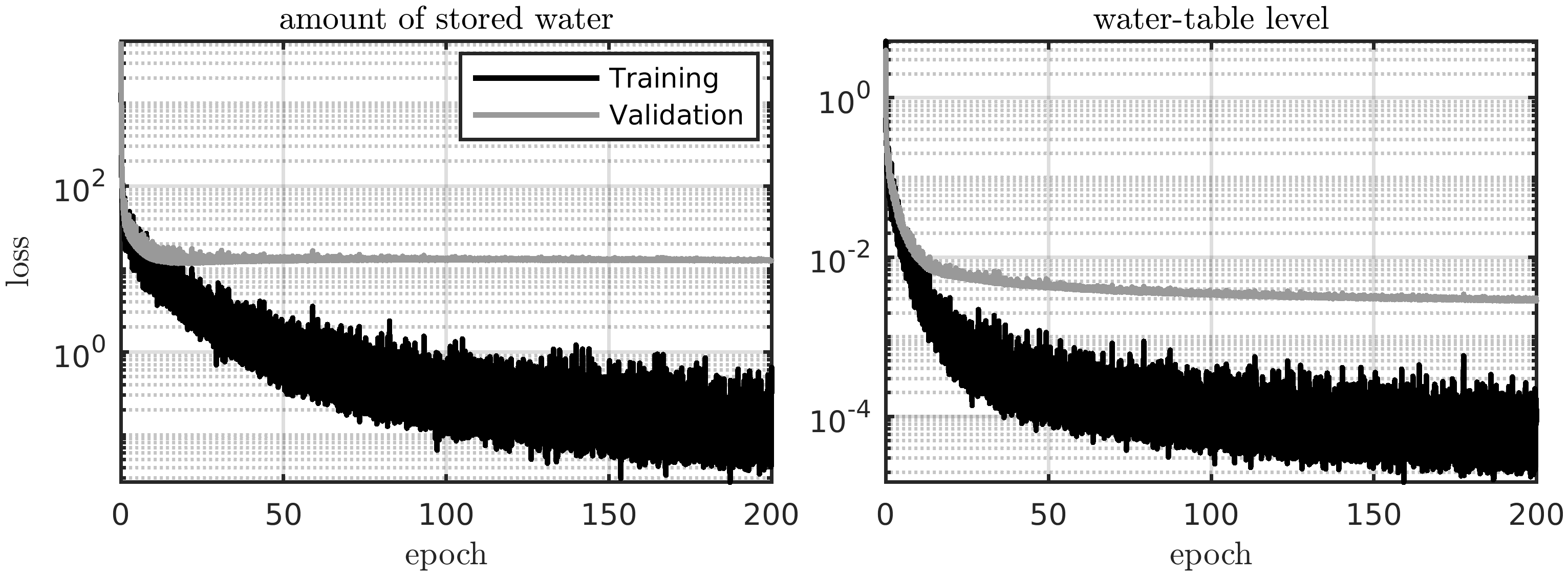}
\caption{Training and validation loss for the amount of stored water (left) and water-table level (right) as a function full training cycles in the stochastic optimization (epoch).} \label{fig:network_train}
\end{figure*}

The value of the loss function for the training and validation data is shown in Figure \ref{fig:network_train}. The losses are shown for both networks, i.e., for prediction of the amount of stored water $V^w$ and another for water-table level $W^l$ prediction. The figures  show that approximately 150 epochs are needed to reach the maximum generalization capability for both components. 

We applied the trained network to predict the amount of stored groundwater and water-table level from images of test data, comprising of 3000 different basement surface profiles and physical parameters as explained previously. 
In addition, we have applied the trained network to the additional 200 samples with two jumps in the basement profile.
One must note that the proposed neural networks-based approach enables direct estimates of the amount of stored groundwater in a heterogeneous porous material rather than separately estimating the porosity field and the aquifer geometry. Both the porosity field and the geometry could potentially have many unknowns, which increase the overall computational demand. 

\begin{figure*}
\begin{center}
\includegraphics[width=\textwidth]{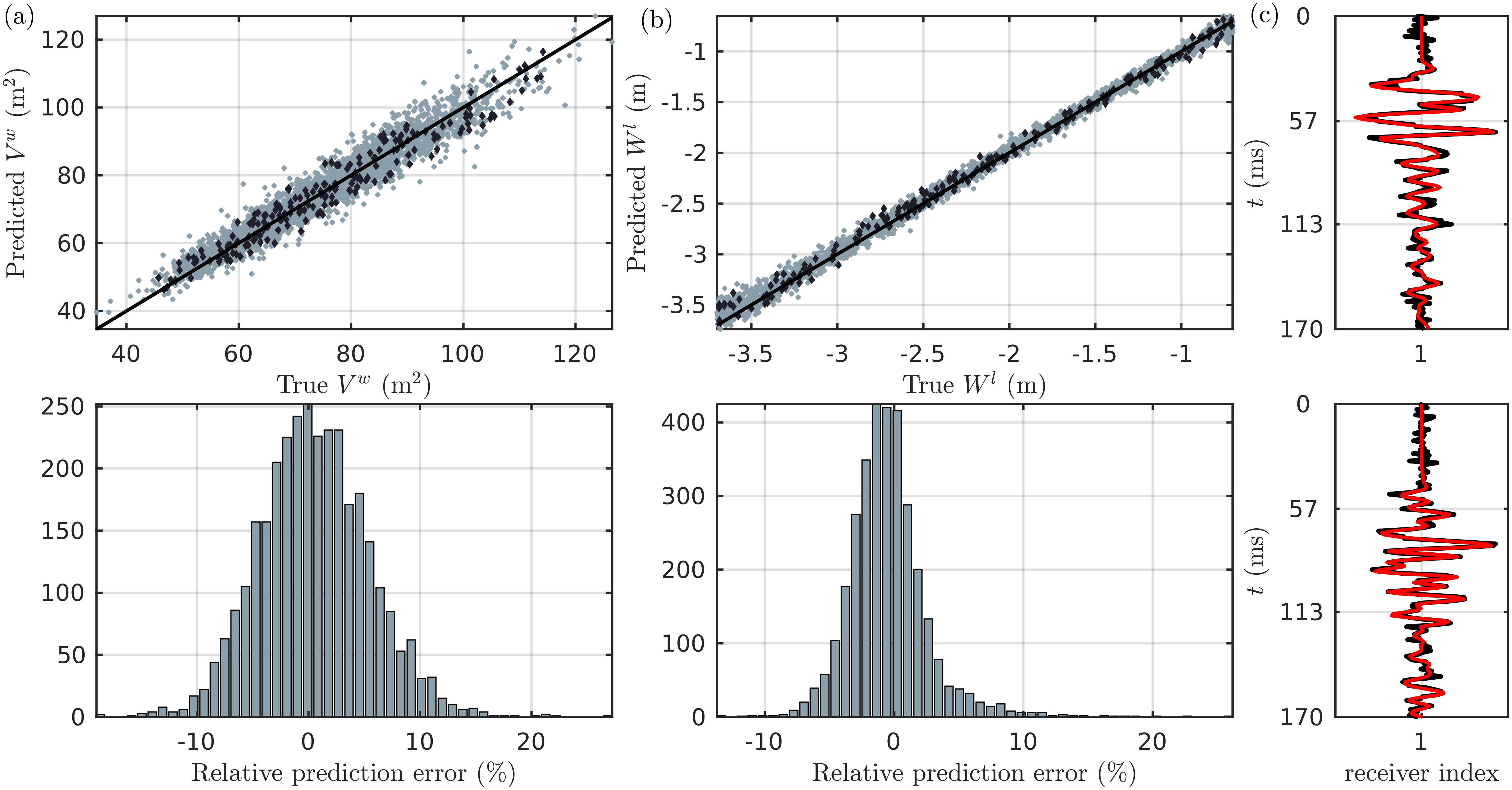}
\end{center}
\caption{Predicted amounts of stored water (a), water-table levels (b), and two example signals for noisy (black) and noise-free (red) data for the test data with $A = 0.011$ and $B = 0.248$ (c). Bottom row shows histograms of the relative prediction error (difference between the predicted and true values). Black diamond symbols on the top pictures show the samples that have two jumps in the basement profile.}
  \label{fig:results_p1}
\end{figure*}

\begin{figure*}
\begin{center}
\includegraphics[width=\textwidth]{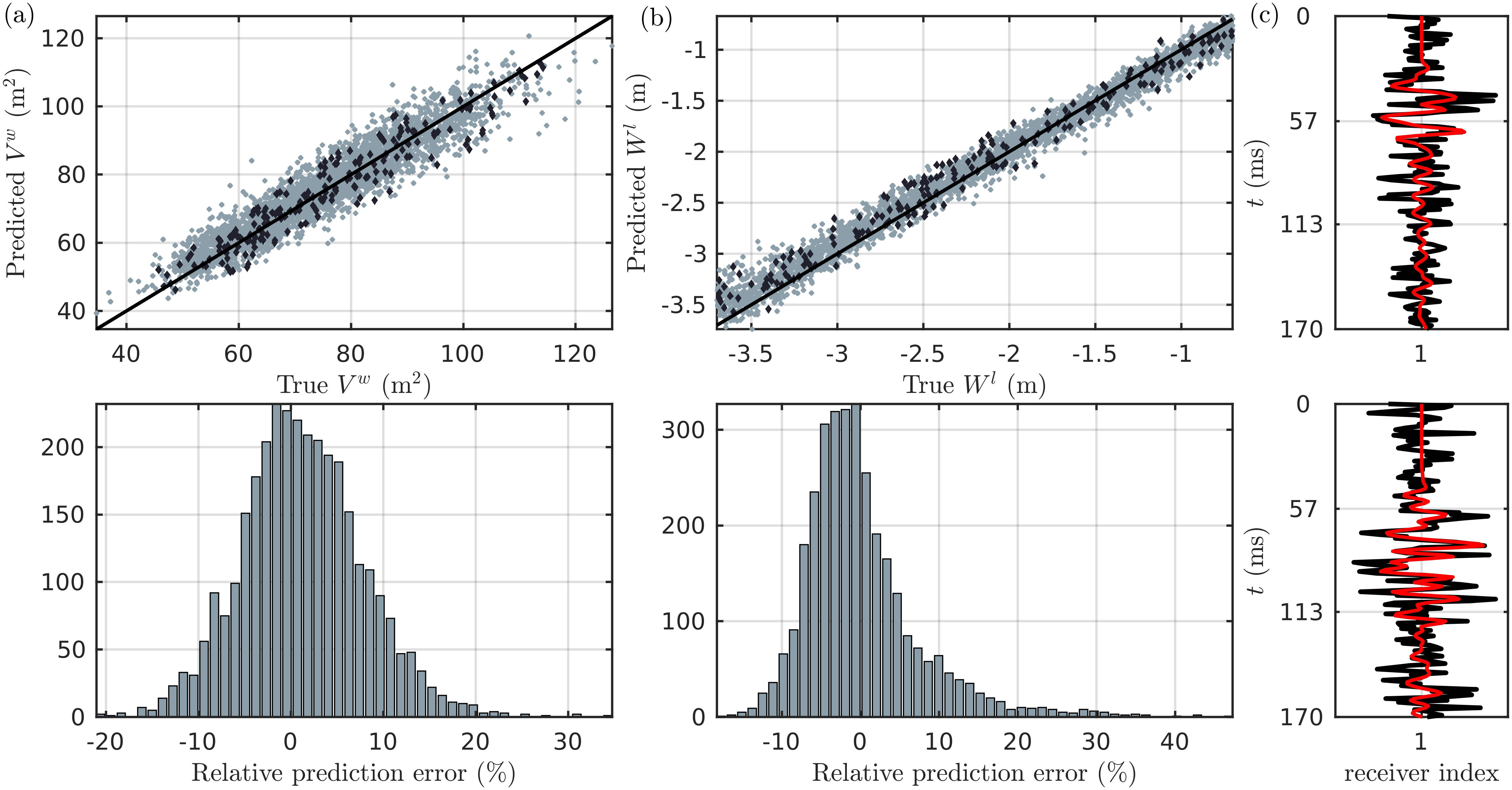}
\end{center}
\caption{Results with noise parameters of $A = 0.045$ and $B = 0.497$. Otherwise same caption as in Figure \ref{fig:results_p1}. }
  \label{fig:results_p2}
\end{figure*}

Figure \ref{fig:results_p1} shows estimates for the test data, contaminated with the white noise component of a moderate level, and Figure \ref{fig:results_p2} shows results for the high noise level. In both cases, the noise level for the amplitude dependent part is assumed to be very high. The figures also include relative prediction error histograms. Clearly, the network yields accurate predictions for both estimated aquifer components. As expected, the error histograms are slightly wider for the higher noise level.

We want to study how sensitive the network is to noise and hence evaluate the network with various levels of noise. As an indicator of accuracy, we use the normalized root-mean-squared error (NRMSE), which is defined as
\begin{equation}
\label{eq:nrmse}
\text{NRMSE}_{\Theta}\ (\%) = 100\frac{\sqrt{T^{-1}\sum_{\ell=1}^{T}\left(\Theta^{\mathrm{predicted}}_{\ell}-\Theta^{\mathrm{true}}_{\ell}\right)^2}}{\Theta^{\mathrm{true}}_{\max}-\Theta^{\mathrm{true}}_{\min}}, 
\end{equation}
where $\Theta$ is either the amount of stored water $V^w$ or water-table level $W^l$ and $T$ denotes the number of samples in the test dataset.

\begin{figure*}
\begin{center}
\includegraphics[width=\textwidth]{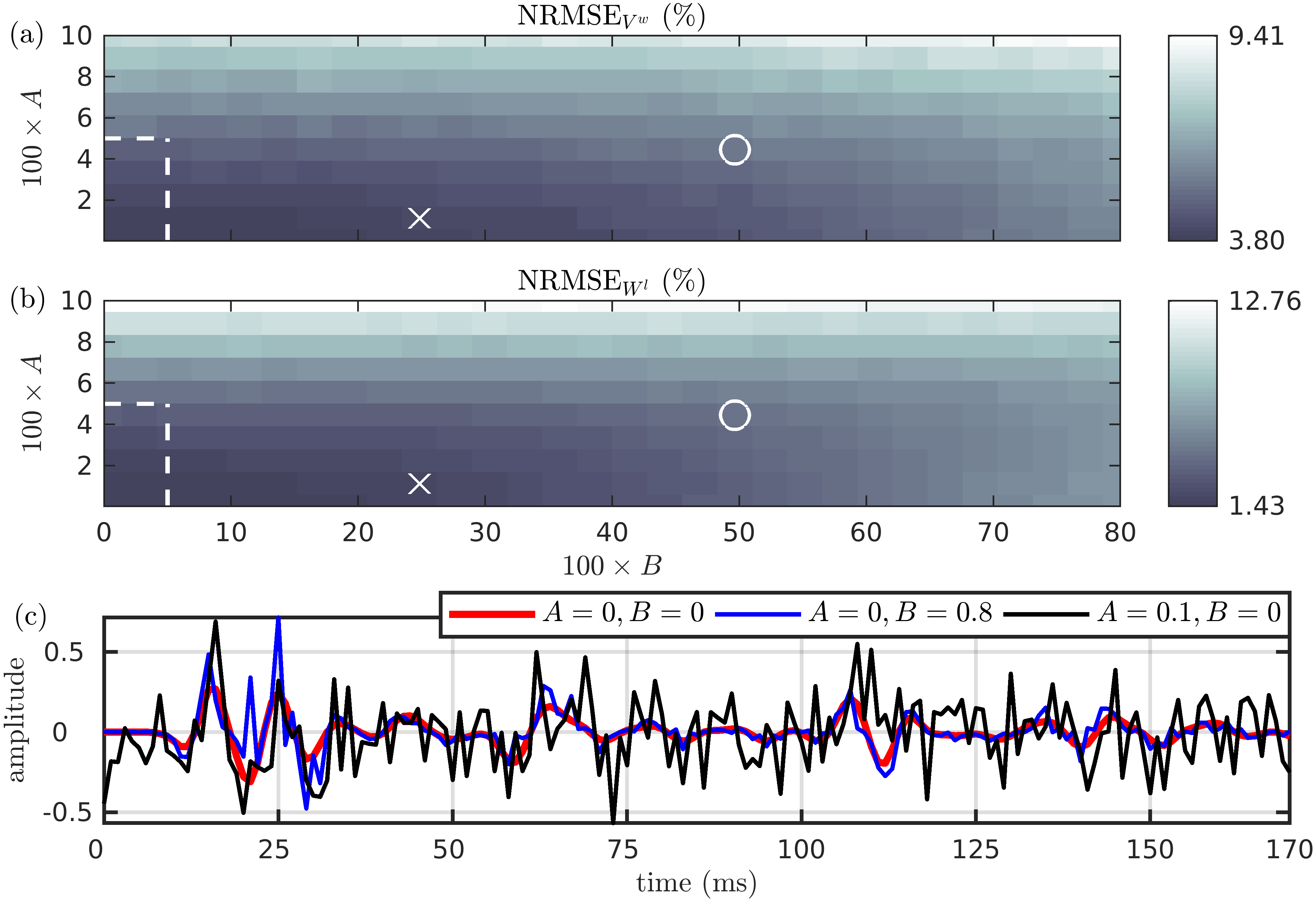}
\end{center}
\caption{Normalized RMSE as a function of noise parameters $A$ and $B$ for the amount of water stored in the aquifer $V^w$ (a) and the water-table level $W^l$ (b). The dashed white box in the bottom-left corner shows the parameter space used in the training data. In addition, the white '$\times$' data are visualized in Figure \ref{fig:results_p1} and the white '$\circ$' in Figure \ref{fig:results_p2}. An example data signal as a function of time with different choices of $A$ and $B$ is shown in (c).}
  \label{fig:results_mat}
\end{figure*}

Figure \ref{fig:results_mat} shows a NRMSE as a function of the parameters $A$ and $B$ for both estimated parameters $V^w$ (a) and $W^l$ (b) and one source/receiver pair signal with different values of parameters $A$ and $B$ (c).  Results indicate that the estimation is more sensitive to stationary white noise ($A$) than to the relative noise ($B$), suggesting that the signal arrival times are more important than the amplitude.  Noise realizations show that when the parameter $A$ is set to 0.1 (and $B=0$), a major part of the signal is swamped by the noise which leads to inaccurate estimates. With $A=0$ and $B=0.8$ the noise-free signals structure is clearly identifiable. In Figures \ref{fig:results_mat}a and \ref{fig:results_mat}b, the selections of $A$ and $B$, used in Figures \ref{fig:results_p1} and \ref{fig:results_p2} are highlighted with white '$\times$' and '$\circ$' markers, respectively. Furthermore, the upper bounds for the noise parameters used in the training are highlighted with a dashed white line. 

\section{Discussion}

The results suggest that a convolutional neural network can provide an efficient approach to estimate the water stored in an aquifer and the water-table level from seismic data. The accuracy of the neural network depends on how well the training data reflects the real situation. In practice, this means that the prior model used in the training data should be general enough to support the expected scenarios found in the test data. In the studied case, the network produces accurate estimates for the primary unknowns also for test samples in which the basement profile was assumed to be more complex than the prior model assumes in the training phase.  Nevertheless, a {\it priori} information, i.e., possible estimates from the earlier survey studies or drilling samples, is very useful. Without a {\it priori} information, the model used to generate samples would need to be very general and would potentially become too computationally demanding to be practical.
  
In this study, we focused on an aquifer located in the Virttaankangas water supply facility where we have adequate a {\it priori} information. Based on earlier studies \cite{artimo, maries16}, the structure of the air- and water-saturated zones and bedrock layer is justified. We estimated the water stored in the aquifer and water-table level while the basement profile and all other material parameters were considered as nuisance (uninteresting) variables/parameters. This marginalization allows us to deal with the cases in which material parameters and basement surface profile are changed by seasonal fluctuations and/or earthquakes, leading to a potential tool for monitoring purposes.

In our case, the ground surface together with the sensor setup is assumed to be known accurately and also that they remain unchanged during time. This setup is reasonable if the purpose is to monitor water levels (time-lapse measurements) in a chosen location (e.g. monitoring over-extraction of groundwater for warning). In our framework, the model trained using (simulated) data for a fixed location may not work for another setup without re-training using data corresponding to the new setup. A more general model can be trained by varying also the surface profile and parameter fields more extensively during the generation of the training sets. However, this probably requires larger training sets. Uncertainties related to sensor setup and ground surface are potential future research topics, especially in the 3D test cases.

This study is performed with synthetic data, but the future goal is to combine synthetic and real datasets. At first, the real data could be used as the test data to provide a more realistic assessment of the framework. In addition, if a suitable amount of real data becomes available, the model trained by simulations can be fine-tuned by using the real data, e.g., by replacing the training set with the real data and re-starting the training procedure.  The use of real data can potentially require additional elements to the physical model. These additions include the anisotropy \cite{puente08, Shukla} and more comprehensive attenuation model \cite{8492340}. Also, when extending the work to real measurements, we may need to reconsider the network architecture (possibly extend, e.g., the number of layers and filter sizes) as well as physical parameter ranges.

\section{Conclusions}\label{sec:conclusions}

In this study, we continued the investigations begun in \cite{lahivaara14, lahivaara15, lahivaara_malmo}, where the general aim was to quantify aquifer properties from passive and active seismic signals using full waveform inversion of a coupled poroelastic-elastic model.  Here, we focused on the problem of estimating the actual water stored in an aquifer and the water-table level for a 2D case study for which the material parameters were modelled as random Markov fields, and used a CNN to marginalize the uncertainty over the unknown material parameters. Synthetic data was generated by a coupled poroviscoelastic-elastic model and corrupted by a two component random noise. An inverse crime was avoided by using an accurate model to generate the test data, while less accurate model was used when generating the training data. We generated an additional 200 test samples, in which the basement profiles are more complex than the prior model assumes.

Results of this study support that the CNN can be used to estimate the aquifer quantities of interest with a wide variety of noise levels, while nuisance parameters can successfully be marginalized. The error histograms for both stored water and water-table show promising accuracy in terms of relative prediction error and bias. The proposed approach yields feasible estimates, with a very substantial reduction in the computational time as compared to what it would have taken from an accurate wave equation solver-based model to estimate the material field in the aquifer geometry.

The aim of the study was not to model a real situation, which would require a 3D subsurface model, but to establish feasibility. The focus was on building a monitoring tool for a known aquifer with material uncertainty. For the 2D model studied in this work we were able to marginalize over the physical parameters used to define the subsurface properties and obtain feasible estimates of the quantities of interest using the CNN.

\section{Acknowledgments}

This work has been supported by the strategic funding of the University of Eastern Finland and by the Academy of Finland (Finnish Centre of Excellence of Inverse Modelling and Imaging). The authors wish also to acknowledge CSC - IT Center for Science, Finland, for computational resources. A contribution also from Trust-GeoInfra (Formas project 2012-2017) supporting Uppsala University contribution in this study. We thank the editor and two anonymous reviewers for their constructive comments that helped improved the quality of our article.

\end{document}